\documentclass[%
 reprint,
 amsmath,amssymb,
 aps,
]{revtex4-2}

\usepackage{graphicx}
\usepackage{dcolumn}
\usepackage{bm}
\usepackage{aas_macros}
\usepackage{xcolor}

\begin{document}

\preprint{APS/123-QED}

\title{From known to unknown: cosmic rays transition from the Sun, the Galaxy, and the Extra-Galaxy}

\author{Yu-Hua Yao}
\email{yaoyh@cqu.edu.cn}
\affiliation{College of Physics, Chongqing University, 401331, Chongqing, China}
\affiliation{Key Laboratory of Particle Astrophysics, Institute of High Energy Physics, Chinese Academy of Sciences, 100049, Beijing, China
}

\author{Yi-Qing Guo}
\email{guoyq@ihep.ac.cn}
\affiliation{
Key Laboratory of Particle Astrophysics, Institute of High Energy Physics, Chinese Academy of Sciences, 100049, Beijing, China
}%
\affiliation{
College of Physics, University of Chinese Academy of Sciences, 100049, Beijing, China
}%
\affiliation{Tianfu Cosmic Ray Research Center, 610000, Chengdu, Sichuan, China}

\author{Wei Liu}
\email{liuwei@ihep.ac.cn}
\affiliation{
Key Laboratory of Particle Astrophysics, Institute of High Energy Physics, Chinese Academy of Sciences, 100049, Beijing, China
}%
\affiliation{Tianfu Cosmic Ray Research Center, 610000, Chengdu, Sichuan, China}

\date{\today}

\begin{abstract}
The Sun stands out as the closest and clearest astrophysical accelerator of cosmic rays, while other objects within and beyond the galaxy remain enigmatic. It is probable that the cosmic ray spectrum and mass components from these celestial sources share similarities, offering a novel approach to study their origin. In this study, we analyze of spectra and mass in the energy range from MeV to 10~EeV. We find: (1) the  mean- logarithmicarithmic mass $\rm\left\langle lnA \right\rangle$ distribution with energy exhibits much clearer feature structures than the spectra; (2) a 100~TeV bump is presented in the $\rm\left\langle lnA \right\rangle$ distribution; (3) for protons, the knee is located at $\sim2$ PeV, the boundary between the galaxy and extra-galaxy occurs at $\sim30$ PeV, marked by a sharp dip; (4) the all-particle spectrum exhibits hardening at $\sim30$~PeV due to the contribution of nearby galaxies, and the extra-galactic dominate $\sim0.7$~EeV. We hope the LHAASO experiment can perform spectral measurements of individual species to validate our results. 
\end{abstract}

\maketitle


\section{Introduction} 

The discovery of the cosmic ray (CR) PeVatron represents a significant milestone in understanding the origin of CRs \citep{2021Natur.594...33C}, which is among the most pressing frontier scientific issues in modern astrophysics. Previously, it was widely believed that CRs with energies below the all-particle spectrum structure "knee" (around 1~PeV) originated from celestial bodies within the Milky Way \citep{kulikov1959size}. However, the Cygnus bubble suggests that the Milky Way is also capable of accelerating CRs beyond 10~PeV \citep{2023arXiv231010100L}. The crucial question raised is the location of the maximum acceleration limits of galactic sources. Energy spectra and mass composition are the most direct observables for solving this puzzle. 

CRs originate from three spatial scales: the Sun, the galaxy, and the extra-galaxy. The Sun, being the closest celestial body, allows for observations and analysis to understand the physical processes within it. Compared to other celestial bodies, we have amassed the most experimental observation data on the Sun, providing researchers with a clear understanding of the temporal and spatial characteristics of each eruption event, the study of energy spectrum and features of high-energy events, and changes in composition and surrounding environments. The Sun has the capacity to accelerate ions to energies ranging from MeV to a few GeV. The energy spectrum of a solar energetic particle (SEP) is typically fitted as a power-law with an exponential \citep{2006GMS...165..103V}, with the spectral 'knee' occurring at around 1~GeV. Observations suggest that the spectrum of lighter elements begins to roll off at a lower energy per nucleon during the spectral steepening \citep{reames1999particle}. To date, the Sun is the most thoroughly studied object capable of accelerating CRs.

The galaxy is considered the next level for accelerating CRs beyond the solar system. Satellite experiments conduct fine spectral measurements of different species of CRs, revealing several observed fine structures, including the primary proton and helium hardening at around 200 GV, and a further bump at 10 TV \citep{2017ApJ...839....5Y,2018JETPL.108....5A,2019SciA....5.3793A,2021PhRvL.126t1102A}. The 10~TeV spectral steepening, combined with anisotropy at around 100~TeV \citep{2005ApJ...626L..29A,2009ApJ...692L.130A}, is considered the signature of a nearby hidden CR accelerator \citep{2019JCAP...10..010L,2021PhRvD.104j3013F}. Galactic CR spectra also exhibit the characteristics of a knee and a lighter cutoff, followed by heavier elements. However, with increasing energy, lower CR fluxes make detection more challenging and lead to a lesser understanding of them. The knee region and below presents some unresolved issues.

It is generally believed that extra-galactic CRs begin to dominate as we move to higher energy levels. However, the question of where extra-galactic CRs begin to dominate is still a matter of debate. Two spectral features, commonly referred to as the second knee at approximately 500~PeV \citep{2005PhLB..612..147B}, or the ankle at approximately 3~EeV \citep{2005A&A...443L..29A,2015PhRvD..92b1302G}, are conventionally understood as points of transition from galactic to extra-galactic. However, explanations are highly dependent on the model \citep{2019PrPNP.10903710K}. Moreover, determining the mass number of individual CR primaries is very difficult due to large fluctuations; only the mean mass number A or a few groups of elements of the CR flux can be obtained. Therefore, the precise intricacies remain a subject of contention, leaving room for further measurements or the development of novel methods.

Given the extensive research on solar CRs and the detailed observations of the transition from the solar to the galactic systems, this study aims to investigate the transition from the galactic system to the extra-galactic system. By analyzing the energy spectrum and mass composition of particles ranging from MeV up to EeV, we aim to identify similarities between these transitions. Utilizing a wealth of datasets, we utilize a phenomenological function to describe the spectra of each individual element, as detailed in Section 2. Subsequently, in Section 3, we present the results related to the all-particle spectrum and mean logarithmic mass, emphasizing the common trends identified between these two transitions. Finally, Section 4 offers a concise summary of our findings.

\section{Component Morphology and Method}
\begin{figure}[!ht]
\centering
\includegraphics[width=0.49\textwidth]{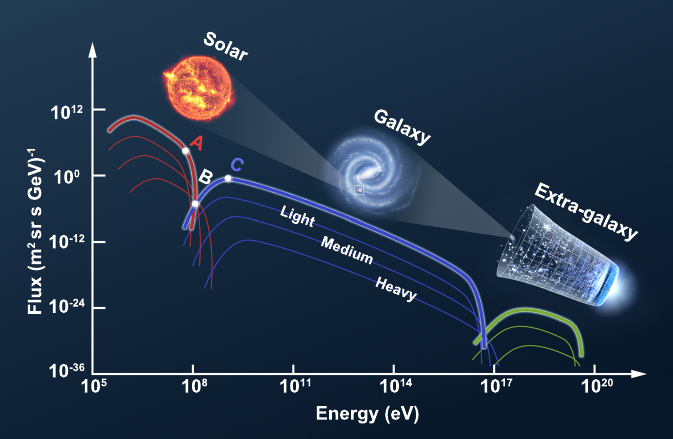}
\caption{ The cartoon of spectral feather of charged particles from solar system, to the galactic, and to the extra-galactic. The spectra of each nucleus in solar energetic events exhibit a maximum energy cut-off, with lighter nuclei dropping first, followed by intermediate ones, and ultimately iron-dominated nuclei. At higher energies, lighter galactic CRs begin to dominate. Point A presents the energy where protons start to drop, point B stands for the interaction where solar protons and galactic iron intersect, and the location dominated by galactic protons as point C.
}
\label{fig:cartoon}
\end{figure}

As mentioned earlier, solar CRs have been extensively studied, making them an ideal standard for investigating the relatively less understood GCRs and the transition from the galactic to the extra-galactic.
As depicted in Figure \ref{fig:cartoon}, the spectra of each nucleus in SEP exhibit a maximum energy cut-off, with lighter nuclei dropping first, followed by intermediate ones, and ultimately iron-dominated nuclei. At higher energies, lighter galactic CRs begin to dominate. In other words, the signature of the transition in composition is the gradual disappearance of solar elements from proton to iron and the increase of a lighter or intermediate galactic component. We designate the energy where protons start to drop as point A, the interaction where solar protons and galactic iron intersect as point B, and the location dominated by galactic protons as point C. That means the mean mass with energy plots should start increasing at A, reaching the peak at point B, and falling back to a trough at point C. It is natural to assume and investigate whether there are similar structures and changes at the transition from within the Milky Way to outside the Milky Way.


The observations for particles with energies ranging from MeV to 10~EeV are utilized as probes for following analysis. To be specific, two significant SEP events, occurring on 29 October 2003 and 25 August 1998 \citep{2008AGUSMSH41A..10M,2007sdeh.book..303M}, have been chosen for analysis, providing composition and energy spectra for key elements from hydrogen to iron with high statistical accuracy. For CRs outside the solar system, we utilize data from space satellite experiments such as AMS02, DAMPE, CREAM, etc., to obtain the nucleon spectrum. At higher energies, partial available nuclei (H, He, Fe) spectra, all-particle spectrum, as well as mean  logarithmic mass distributions are adopted. 

To achieve a consistent description, the energy spectra of individual elements (P, He, C, N, O, Mg, Si, Fe) within the MeV to 10~EeV range are described using a phenomenological function, i.e., multiple power-law segments. A cut-off power-law function is employed to capture the spectral features of solar energetic events, with three segments for galactic CRs and an additional two segments for the extra-galactic components. Each elemental spectrum is individually fitted with this complementary function. Mathematically, the overall fitting function is described by 
\begin{equation}\label{eq:phi}
\begin{split}
    F &=\Phi_{pl,1} (Solar) \\
       &+ \Phi_{sbpl}+\Phi_{pl,2} + \Phi_{pl,3} (GCR)  \\
       &+ \Phi_{pl,4} + \Phi_{pl,5} (EGCR)
\end{split}
\end{equation}
with 
\begin{equation}\label{eq:pl}
\Phi_{pl}(E)=\Phi_i~E^{\gamma_i} e^{-\frac{E}{E_{c,i}}},
\end{equation}

\begin{equation}\label{eq:sbpl}
    \Phi_{sbpl}(E) = \Phi_0  E ^ {-\alpha_1}  \left[ 1 + \left( \frac{E}{E_{b,0}}\right)^{1 / \delta} \right] ^{(\alpha_1 - \alpha_2) \delta} e^{-\frac{E}{E_{c,0}}}. 
\end{equation}

The fitting result for the proton is depicted in the left panel of Figure \ref{fig:fit}, while the fitting for all key species are shown in different colors on the right. The sum of all fittings represents the all-particle spectrum, which is compared with experimental observations and also presented in Figure \ref{fig:fit}. The fitted parameters of each species are listed in Table \ref{tab:para}. A hyperbolic cosine is adopted for the solar modulation which is approximately 450 MV, and a similar galactic modulation as 0.1~EeV in the data comparison. 

The mean logarithmic mass, an often-used quantity to characterize the
CR mass composition, below PeV could be calculated with fitting function as well as direct observational spectra data of each element, using 
\begin{equation}\label{eq:logA}
\langle\ln \text A\rangle=\frac{\sum \ln A_i \times \text {F}_{\mathbf{i}}}{\sum \text {F}_{\mathrm{i}}}
\end{equation}
The trends could be compared with $\rm\left\langle lnA \right\rangle$ observations from ground-based facilities.

\begin{table*}[!htb]
\caption{Spectral parameter of different species.}\label{tab:para}%
\begin{tabular*}{\textwidth}{@{\extracolsep\fill}c|cc|cccccccc|cccc}
\hline
 & \multicolumn{2}{c|}{Solar} & \multicolumn{8}{c|}{Galactic}  & \multicolumn{4}{c}{Extra-galactic}  \\
& $ \Phi_1 $  & $ \gamma_1$  &  $\Phi_0$ &  $\alpha_1$     &  $\delta$ & $\alpha_2$ & $\Phi_2$ & $\gamma_2$ & $\Phi_3$ & $\gamma_3$   &  $ \Phi_4 $  & $ \gamma_4$ & $ \Phi_5$ &  $\gamma_5$  \\
 \hline
P & $1\times1.5^{6}$ &0.87&$1.16\times10^{5}$&-3.01&1.19&3.05&50&2.12&300&2.4&12&2.4&15&2.5  \\  
He& $2\times10^{5}$ & 0.6 & $1.18\times10^{4}$ & -3.32&1.27&2.93&60&2.16&43&2.26&0.3&2.3&0.04&2.26 \\
C& 1000 & 0.7 & 47.4&-4.07&1.90&3.26&8.10&2.16&5 &2.26&0.01&2.3&0.002&2.26 \\
N& 90 &1&12.14&-3.89&1.9&3.44&2.27&2.16&0.1&2.26&0.001&2.3&$1\times10^{-5}$&2.26 \\
O& 2000 &0.7&9.84&-4.03&1.91&3.24&13.70&2.16&3&2.26&0.05&-2.3&0.003&2.26 \\
Mg& 500 &0.7&0.26&-4.08&1.91&3.25&4.53&2.16&2&2.26&0.1&2.3&$1\times10^{-4}$&2.26 \\
Si& 400 &0.6&0.06&-4.15&1.91&3.18&5&2.16&3&2.26&0.001&2.3&$1\times10^{-4}$&2.26 \\
Fe& 300 &0.7&$8.8\times10^{-4}$&-4.29& 1.9 & 3.05&4.09&2.1&0.05&2&0.01&2.4&$1\times10^{-4}$&2.3 \\
 \hline
\end{tabular*}
\begin{itemize}
    \item  $\rm E_{c}=2.5\times10^{4}Z$~GeV and $\rm 4.5\times10^{6}Z$~GeV for galactic and extra-galactic CRs, Z is the charge of each species. 
\end{itemize}
\end{table*}

To gain deeper insights into the intricacies of the energy spectrum and $\rm\left\langle lnA \right\rangle$, changes are tracked and calculated alongside the energy using the equation:

\begin{equation}\label{eq:cos}
    \Delta=\frac{\Delta y}{\Delta r} = \frac{\Delta y}{\sqrt{(\Delta x)^2+(\Delta y)^2}}.
\end{equation}

This equation effectively captures the variations depicted in Figure \ref{fig:cartoon}, where points A, B, and C are annotated on the energy spectra, corresponding to changes in the mass component. It is important to note that in this context, the change in the energy spectrum is the value after being subjected to a double logarithm, while for $\rm\left\langle lnA \right\rangle$, it is the value obtained from the logarithm and linear.

\section{Results} 
\begin{figure}[h]
\centering
\includegraphics[width=0.5\textwidth]{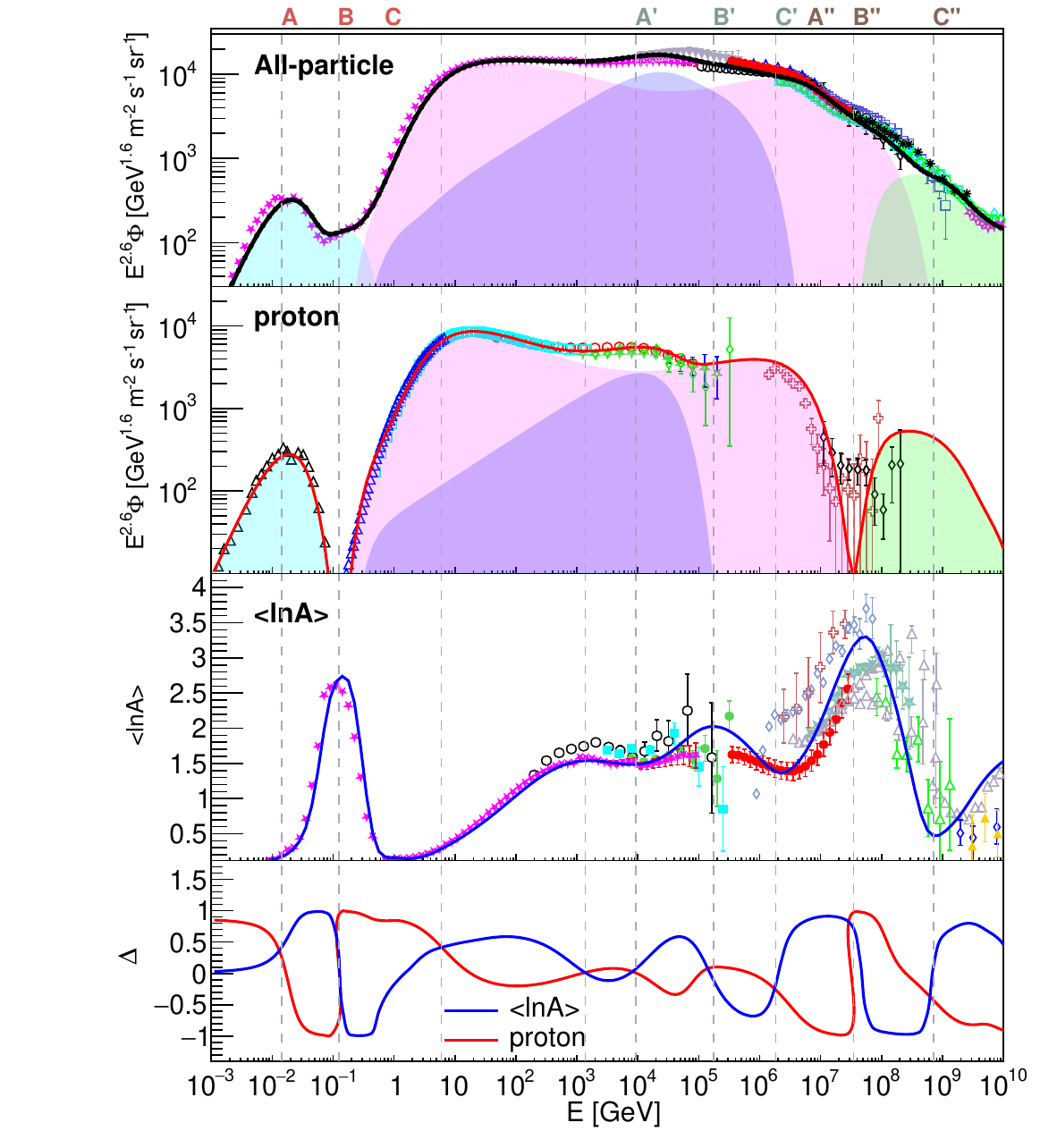}
\caption{ The intersections of the changes in the proton energy spectrum and the distribution of logA are presented as gray dashed lines running through four plots, as detailed in the main text. Three intersections (A-B-C, A'-B'-C', and A''-B''-C'') can determine a transition of the system. Top: All-particle spectrum. The black solid line represents the sum of the fitting function for several key species. The fitting parameters and results of individual elements are listed in Tab.\ref{tab:para} and shown in Fig.\ref{fig:fit}. The light blue shade represents the contribution from SEPs, the pink shade represents the background CRs, the light violet shade comes from nearby sources, and the green shade represents the contribution from extra-galactic sources. The magenta pentagrams represent the total-error-weighted average observational spectra of key species. The data used to calculate the weighted spectra and all-spectra data are also listed in the caption of Fig.\ref{fig:fit}. 
Second: Proton spectrum. The different fitting components are shown on the left of Fig.\ref{fig:fit}, and experimental data comparisons can also be found there. 
Third: $\rm\left\langle lnA \right\rangle$ distribution. The blue solid line is obtained by Eq.\ref{eq:logA} with fitting functions of elemental flux. The magenta pentagrams represent observational spectra of key species from space-born experiments. Other points are observational data from ATIC \citep{2009BRASP..73..564P}, Auger \citep{PierreAuger:2023kjt}, Hires-MIA \citep{2001APh....16....1A}, KASCADE \citep{2005APh....24....1A}, Tunka-133 \citep{2012NIMPA.692...98B}, LHAASO \citep{LHAASO:2024knt}, IceCube \citep{2019PhRvD.100h2002A}.
Bottom: the $\rm\Delta$ distributions of proton spectrum fitting and $\rm\left\langle lnA \right\rangle$. The overall vertical grey dashed lines are determined by their intersections.
}
\label{fig:all}
\end{figure}

Figure \ref{fig:all} presents the spectra and $\rm\left\langle lnA \right\rangle$, and their change vary with energies from MeV to 10~EeV. The first and second pad display the flux of all particles and protons, respectively. It can be seen that function described flux comparison with the observed spectrum from experiments, demonstrating fitting function well align with the experiment data. In both spectral plots, distinct colored shaded areas are utilized to delineate the entire energy spectrum structure. As previously mentioned, the CRs within the solar system are accurately represented by a truncated power-law spectrum, illustrated by the blue shaded area in the plot. The pink shaded area signifies the background spectrum of CRs within the galaxy, while the purple shaded area represents the contribution from nearby sources. Additionally, the green shaded area corresponds to the contribution from extra-galactic flux.

The third panel presents the mean logarithmic mass $\rm\left\langle lnA \right\rangle$ with energy, calculated using Eq. \ref{eq:logA}, and includes observations from ground-based detectors, including the latest data release from LHAASO \citep{LHAASO:2024knt}. It can be seen two significant peaks arise at around 0.1~GeV and 30~PeV, which indicate where the proton ratio is highest in the solar system and within the Milky Way, and one shallow peak presenting at near 200~TeV, which suggest the location of nearby source contributions. The features in $\rm\left\langle lnA \right\rangle$ distribution provides more evidence than that in the spectral plot.

To illustrate the variability of spectra and $\rm\left\langle lnA \right\rangle$ with energy, the changes in the spectra and $\rm\left\langle lnA \right\rangle$ distributions are presented at the bottom of Figure \ref{fig:all}. The intersections of the $\Delta$ distributions of proton spectra and $\rm\left\langle lnA \right\rangle$ are labeled and drawn with vertical lines across the entire canvas. A joint analysis of the intersections, spectral distribution, and the $\rm\left\langle lnA \right\rangle$ could provide a better understanding.

It is evident that three cross-points effectively describe a transition from the solar system to the galactic system, labeled as A-B-C. The proton ratio is highest at point A within the solar system. This is evident from the energy spectrum and $\rm\left\langle lnA \right\rangle$ distribution, where point A represents the peak flux in the energy spectrum and is located at the base of the mountain in the $\rm\left\langle lnA \right\rangle$ distribution. Point B, on the other hand, represents the opposite of point A, where the solar proton proportion decreases to the least and $\rm\left\langle lnA \right\rangle$ reaches the peak. From A to B, the flux of SEP is successively truncated from hydrogen to iron. As from B to C, the CR flux from the galaxy begins to increase, resulting in a transition of $\rm\left\langle lnA \right\rangle$ from being dominated by heavy nuclei to being dominated by light nuclei, where point C has the local maximum proton ratio. This transition occurred from the solar system to the galactic system, where the mass components and spectra are clearly known.


It is worth noting that two other similar transitions occur at A'-B'-C' and A''-B''-C'' regions. The A'-B'-C' region indicates a relatively gradual process of $\rm\left\langle lnA \right\rangle$ transitioning from light to heavy and then back to light, suggesting that nearby source(s) in the Galaxy start to contribute at around several tens of TeV, the nearby iron reaches its peak at around 200~TeV, and the galactic proton flux cuts off at about 2~PeV. The A''-B''-C'' region likely marks the transition from the galactic to the extra-galactic at around 30 PeV, also implying that the dominant flux contributions are from nearby galaxies at energies around 0.7~EV and above.

\section{Summary}

This study introduces a novel method for investigating the transition from the galaxy to the extra-galaxy, as well as the presence of local sources. We conducted a combined analysis of CR energy spectrum and mass composition with energies ranging from MeV to 10~EeV, defined a new variable to characterize the evolving structure, and identified three key points where intersections of the energy spectrum and $\rm\left\langle lnA \right\rangle$ could indicate a system transition. Since the transition of solar CR to galactic is evident, it serves as the standard for understanding nearby sources and the transition to the extra-galactic. Through this work, we found that the distribution of mass composition, $\rm\left\langle lnA \right\rangle$, exhibits clearer trends compared to the energy spectrum. Each transition presents a peak in the $\rm\left\langle lnA \right\rangle$, thus we expect a peak in the $\rm\left\langle lnA \right\rangle$ value near 200~TeV, likely resulting from contributions of nearby sources. Furthermore, the intersections A'' (with a minimum $\rm\left\langle lnA \right\rangle$) represent the galactic proton knee location at 2~PeV, while the intersections B'' (with a peak $\rm\left\langle lnA \right\rangle$ value) indicate the iron reaching its peak, where the transition occurs from galactic to extra-galactic at $\sim$30~PeV, with extra-galactic particles taking precedence over galactic ones at energies surpassing 0.7~EeV. We eagerly anticipate the latest observations on the 100~TeV-100~PeV energy spectra of individual species from LHAASO to validate our findings.

\begin{acknowledgments}
This work is supported by the National Natural Science Foundation of China (Nos. 12275279, U2031110), the China Postdoctoral Science Foundation (No. 2023M730423).
\end{acknowledgments}

\appendix

\section{Fitting results}\label{secA1}
\setcounter{table}{0} 
\setcounter{figure}{0} 
\renewcommand{\thetable}{A\arabic{table}}
\renewcommand{\thefigure}{A\arabic{figure}}
The fitting results of proton spectrum and all-particle spectra are depicted in Figures \ref{fig:fit}. 

\begin{figure*}[!ht]
\centering
\includegraphics[width=0.49\textwidth]{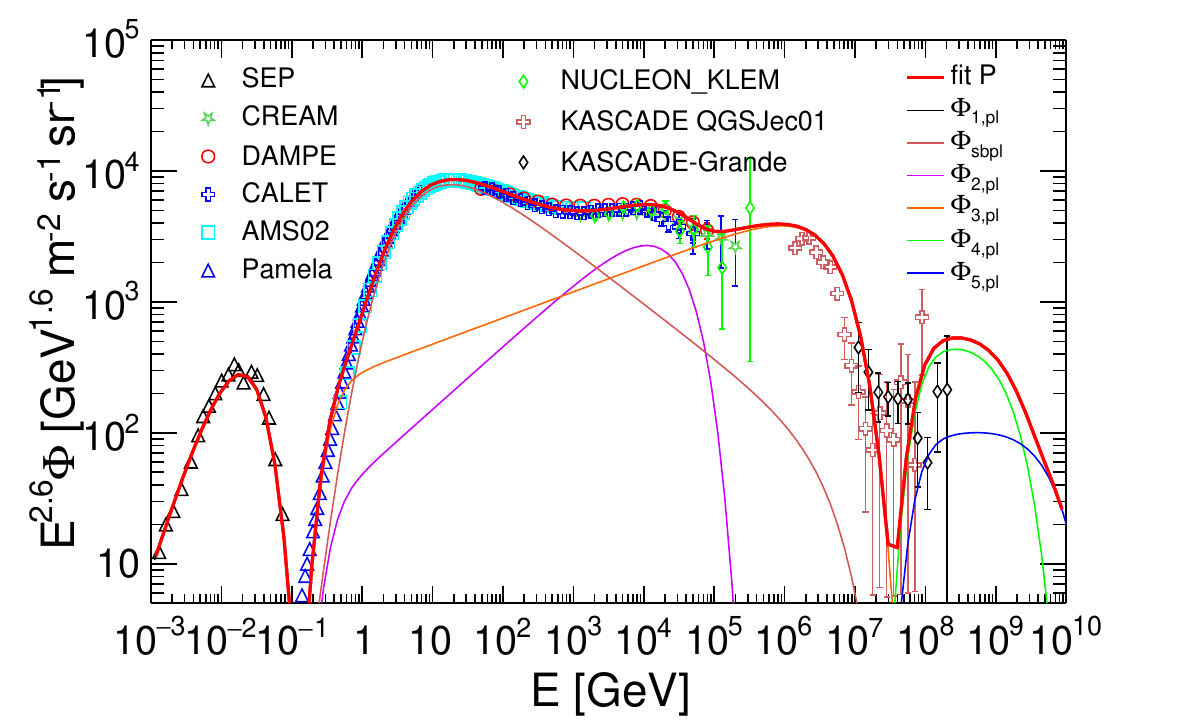}
\includegraphics[width=0.49\textwidth]{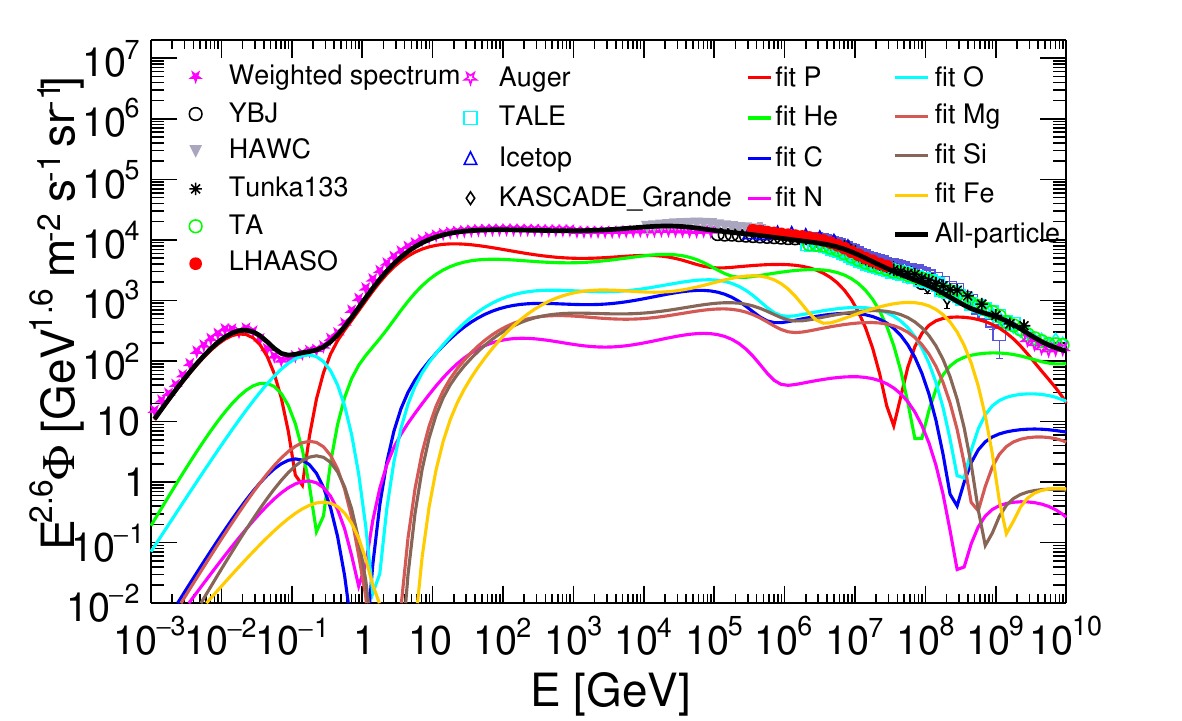}
\caption{Left: Proton spectrum from MeV to EeV, fitted with a multi-segment function and compared with experimental data from experiments \citep{2015PhRvD..91k2017B,2017PhRvD..96l2001A,2020PhRvD.102l2001A,2019PhRvD.100h2002A,2019SciA....5.3793A,2021PhR...894....1A,2017ApJ...839....5Y,2019AdSpR..64.2546G,2013ApJ...765...91A,2013APh....47...54A,2005APh....24....1A,2022PhRvL.129j1102A}. Right: The sum of fitting individual species, compared with all-particle spectrum data from experiments \citep{2020PhRvD.102f2005A,2020APh...11702406B}. The magenta pentagrams represent the total-error-weighted average observational spectra of key species, with data from AMS-02 \citep{2021PhR...894....1A,2021PhRvL.126d1104A}, DAMPE experiments \citep{2019SciA....5.3793A,2021PhRvL.126t1102A}, CREAM \citep{2011ApJ...728..122Y}. Comparing all-particle spectra data are from Tibet-III \citep{2008ApJ...678.1165A}, HAWC \citep{2017PhRvD..96l2001A}, LHAASO \citep{LHAASO:2024knt}, Tunka-133 \citep{2020APh...11702406B}, Auger \cite{2020PhRvD.102f2005A}, TALE \citep{2018ApJ...865...74A}, KASCADE \citep{2005APh....24....1A}, IceTop \citep{2020PhRvD.102l2001A}, KASCADE-Grande \citep{2013APh....47...54A}. Solid lines in different colors represent the fitting results of each nuclei spectrum.
}
\label{fig:fit}
\end{figure*}
\nocite{*}
\bibliographystyle{unsrt_update}
\bibliography{apssamp}

\end{document}